\documentclass[aps,twocolumn,showpacs,prb,floatfix]{revtex4-1}

\usepackage{bm}
\usepackage{graphicx}
\usepackage{amsmath,amssymb}
\usepackage{accents}
\usepackage{float}
\usepackage{wrapfig}
\usepackage{layout}
\usepackage{color}
\usepackage{braket}
\let\%\relax 
\DeclareFontFamily{OT1}{pxr}{}
\DeclareFontShape{OT1}{pxr}{m}{n}{<->pxr}{}
\DeclareSymbolFont{letA}{OT1}{pxr}{m}{n}
\DeclareMathSymbol{\%}{0}{letA}{`\%}
\usepackage{scalerel,stackengine}
\stackMath
\newcommand\reallywidecheck[1]{%
\savestack{\tmpbox}{\stretchto{%
  \scaleto{%
    \scalerel*[\widthof{\ensuremath{#1}}]{\kern-.6pt\bigwedge\kern-.6pt}%
    {\rule[-\textheight/2]{1ex}{\textheight}}
  }{\textheight}%
}{0.5ex}}%
\stackon[1pt]{#1}{\scalebox{-1}{\tmpbox}}%
}
\newcommand{\ii}{\mathrm{i}}
\newcommand{\F}{\mathrm{F}}
\newcommand{\dd}{\mathrm{d}}
\newcommand{\od}{\mathrm{od}}
\newcommand{\BdG}{\mathrm{BdG}}
\newcommand{\imp}{\mathrm{imp}}
\newcommand{\TF}{\mathrm{TF}}

\newcommand{\M}{\mathrm{M}}
\newcommand{\R}{\mathrm{R}}
\newcommand{\A}{\mathrm{A}}

\newcommand{\n}{\mathrm{n}}

\newcommand{\Rot}{\mathrm{rot}}

\newcommand{\Tr}{\mathrm{Tr}}
\def\leftdef{\mathrel{\mathop:}=}

\begin{document}
\preprint{aps/}
\title{Vortex charge and impurity effects based on quasiclassical theory}

\author{Yusuke Masaki}
\email{masaki@vortex.c.u-tokyo.ac.jp}
\affiliation{Department of Physics, The University of Tokyo, Tokyo 113-0033, Japan}
\affiliation{Research and Education Center for Natural Science, Keio University, Yokohama 223-8522, Japan}

\date{\today}
\begin{abstract}
We investigate charge densities around the vortex cores of an s-wave and a chiral p-wave superconductor (SC) in two dimension within quasiclassical theory. 
We consider contributions of particle-hole asymmetry through gradient expansions in first order of a quasiclassical parameter using augmented quasiclassical theory. 
The chiral p-wave SC has two inequivalent vortices: one is charged and the other is uncharged on the basis of Bogoliubov--de Gennes (BdG) equation. We explain this qualitatively distinct charges also by the augmented quasiclassical theory. In addition, we find that much larger charge is induced for the charged vortex of the chiral p-wave SC compared with the vortex of the s-wave SC in the clean system. We also confirm this enhancement through quantitative comparison with results based on the BdG theory. We also study effects of Born-type impurities using self energy formalism and find that the larger charge density is rather suppressed in contrast to the s-wave's one. Furthermore we study the local density of states as another equilibrium property of particle-hole asymmetry. 
\end{abstract}
\maketitle

\section{Introduction}
Properties of vortices in a rather clean superconductor (SC) are dominated by electronic bound states in the vortex core, called Caroli--de Gennes--Matricon(CdGM) mode~\cite{Caroli1964} or Andreev bound states~\cite{andreev1965thermal, PhysRevB.54.13222}. 
For example, the scattering of the bound states dominantly contributes to the viscous flow of the vortex, while the particle-hole asymmetry of the CdGM mode is one of the origins of the Hall effect in the flux flow states~\cite{Larkin1986,Blatter1994,PhysRevB.54.13222,Kopnin200107}. 
The charging effect of the vortex core is another consequence of the particle--hole asymmetry of the CdGM mode. Recently vortex core states have been attracting another great deal of attraction since the vortex of topological SCs such as a two-dimensional (2D) spinless chiral p-wave SC possesses a stable zero energy state called the Majorana state~\cite{RevModPhys.82.3045, RevModPhys.83.1057}. Majorana bound states obey the non-Abelian statistics and enable a fault tolerant quantum computation~\cite{RevModPhys.80.1083}, which will be done by exchanging the vortices adiabatically~\cite{IvanovStephanovichZhmudskii1990}. Deeper understanding  of vortex properties of topological SCs is thus highly required from both fundamental and applicational viewpoints.

The charging effects are mainly caused by particle-hole asymmetry, particularly, in the following two aspects.
One is the finite slope of the density of states in the normal states at the Fermi level. It results in the chemical potential difference between the normal and superconducting states. The charging effects in this aspect were phenomenologically studied by Khomskii and Freimuth\cite{Khomskii1995} and Blatter {\it et al.}\cite{PhysRevLett.77.566}. Recently Ueki {\it et al.} confirmed this mechanism in the quasiclassical theory with quantum correction of order $1/(k_{\F}\xi)$, a quasiclassical parameter denoted by the combination of the Fermi momentum $k_{\F}$ and the coherence length of the superconductivity $\xi$.\cite{Ueki2018} On the other hand, Hayashi {\it et al.} showed that the charging effects occurred even when the normal state was particle-hole symmetric for example in the case of electron gas in two dimensions\cite{Hayashi1998}. They argued using the microscopic Bogoliubov--de Gennes (BdG) equation as follows: the charging effects are due to the difference between the amplitude of the particle-like wave function $|u_{l}(r)|^{2}$ and that of the hole-like wave function $|v_{l}(r)|^{2}$: the quasiparticle state with angular momentum $l$ in the presence of a single vortex is described by the mixture of particle and hole wave functions $(u_{l}(r)e^{\ii (l + 1/2) \theta }, v_{l}(r)e^{\ii (l-1/2)\theta})$ with the cylindrical coordinate $(r, \theta)$. After the work Ref.~[\onlinecite{Hayashi1998}], the vortex-core-charges for the chiral p-wave SCs were calculated by Matsumoto and Heeb by taking the screening effects into account\cite{MatsumotoHeeb2001}. 
The chiral p-wave SCs have chirality, which is an angular momentum of the Cooper pair. Hence there are two non-equivalent single vortices defined by the relative directions between the vorticity and the chirality of the Cooper pair. Andreev bound states of these two vortices exhibit remarkably different properties on impurity effects by Born-type scatters\cite{Kato2000,JPSJ.71.1721,PhysRevLett.102.117003,Kurosawa2015}, nuclear magnetic relaxation rates\cite{Hayashi2003}, chiral optical absorptions\cite{matsumoto1999chiral}, and charging effects\cite{MatsumotoHeeb2001}. In particular, the Anderson's theorem is locally recovered because the phase windings of the quasiparticles owing to the vorticity and chirality around the vortex core center cancel with each other\cite{KatoHayashi2001,JPSJ.71.1721}. As a result, the core of the antiparallel is similar to that of an s-wave SC. This cancelation leads to robustness against the Born-type scatterers. It also leads to the particle-hole symmetry of the low energy CdGM modes and thus the charging effects at the origin are strongly suppressed.

A quasiclassical theory is powerful tool to describe electronic structures in the system with slowly varying potentials such as type-II SCs with large $k_{\F}\xi$, though a conventional Eilenberger equation is automatically particle-hole symmetric~\cite{Eilenberger1968a, larkin1969quasiclassical}. The particle-hole asymmetry inherent to the microscopic electronic structure in the vortex core, which is of order $1/(k_{\F}\xi)$, can be described in quasiclassical theory with a gradient expansion\cite{Ohuchi2017,Ueki2018,Masaki2018charge}. The augmentation of the quasiclassical theory is basically done through the gradient expansion to take account of the quantum correction including particle-hole asymmetry\cite{KadanoffBaym1962,Serene1983,Eckern1981a}, but it should be performed carefully for charged systems. In order to introduce the gauge invariant distribution function, a phase factor was introduced for the Green's function in the normal state by Levanda and Fleurov, otherwise a complicated Gauge transformation remains in the Wigner representation\cite{Levanda1994,Levanda2001,Hansch1983,Mahan1987}. In the superconducting state, Kopnin developed the convenient kinetic theory using a similar phase factor to performing the Wigner transformation~\cite{Kopnin1994} and subsequently
Kita has improved the form of the phase transformation~\cite{PhysRevB.64.054503}.

In the augmented quasiclassical theory, the quantum correction of order $1/(k_{\F}\xi)$ is treated as a perturbation and we retain the gradient expansion up to the first order. Hence it is non-trivial whether the augmented theory can describe behaviors unique to the non-zero quantum corrections. 
In this paper, we apply the augmented quasiclassical theory to the s-wave SC and the chiral p-wave SC~\cite{Mackenzie2003, MaenoKittakaNomuraYonezawaIshida2012}, and study their charging effect. Although they are  accessible also by the BdG theory, the advantage of the quasiclassical theory is that we can study systems with large $k_{\F}\xi$, and effects of randomness using the self energy, and extend to the non-equilibrium phenomena. We demonstrated in Ref.~\onlinecite{Masaki2018charge} that the striking difference between two types of vortices in the chiral p-wave SC, known by BdG theory~\cite{MatsumotoHeeb2001}, was obtained also in this framework. Here we study these strikingly different vortices through the temperature dependence of the vortex charges and the impurity effects on them within the self-consistent Born approximation. We also compare our results with calculations based on the BdG theory. 
The local density of states with asymmetric spectra are studied as another equilibrium property originating from the quantum correction. We demonstrate through these phenomena that effects of discrete spectrum do not appear, and the validity is limited up to the first order of $1/k_{\F}\xi$.

The paper is structured as follows: In Sect.~2, we summarize the formulation to calculate the charging effects in Matsubara frequency formalism. In Sect.~3, we show the numerical results on vortex charges. In Sect.~4, we briefly mention the real frequency formalism and show the asymmetric spectra in LDOS.
In the following part, we use the unit $k_{\mathrm{B}}=1$ in addition to $\hbar = 1$ for convenience.
\section{Formulation} \label{formulation}
In this paper, we consider only equilibrium cases of 2D SCs and charge densities in the Matsubara formalism.
The Cooper pairs  are described by the pair of electrons with spin $\uparrow$ and $\downarrow$ respectively.
We first define the Matsubara Green's functions with Nambu spinors
$\vec{\Psi}(\bm{r},\tau) = {}^{t}[\psi_{\uparrow}(\bm{r},\tau),\psi_{\downarrow}^{\dagger}(\bm{r},\tau)]$ 
and $\vec{\Psi}^{\dagger}(\bm{r},\tau) = [\psi_{\uparrow}^{\dagger}(\bm{r},\tau),\psi_{\downarrow}(\bm{r},\tau)]$  as
\begin{align}
\widehat{G}^{\M} (\bm{r}_{1},\bm{r}_{2};\ii \omega_{n})
&=\begin{bmatrix}
G^{\M}(\bm{r}_{1},\bm{r}_{2};\ii \omega_{n}) & F^{\M}(\bm{r}_{1},\bm{r}_{2};\ii \omega_{n})\\
-\bar{F}^{\M}(\bm{r}_{1},\bm{r}_{2};\ii \omega_{n}) &\bar{G}^{\M}(\bm{r}_{1},\bm{r}_{2};\ii \omega_{n}) 
\end{bmatrix}\nonumber\\
&=\int_{0}^{\beta}\dd \tau e^{\ii \omega_{n}\tau}\widehat{\tau}_{3}
\Braket{\vec{\Psi}(\bm{r}_{1},\tau)\vec{\Psi}^{\dagger}(\bm{r}_{2},0)}.
\end{align}
Here $\omega_{n}$ is the fermion Matsubara frequency at temperature $T$, and $\beta$ is the inverse temperature defined by $\beta = 1/T$.
We use the symbol $\widehat{\cdot}$ to describe a 2 by 2 matrix, and $\widehat{\tau}_{3}$ is the third component of the  2 by 2 Pauli matrices.
We consider the equilibrium case, and Green's function is no longer dependent upon the center-of-mass imaginary-time, while it depends on the center-of-mass coordinate $\bm{R} = (\bm{r}_{1}+\bm{r}_{2})/2$ for inhomogeneous systems. The phase transformation of the Green's function in real space is convenient to take account of the gauge invariance in the mixed representation. 

First of all we formulate the augmented quasiclassical theory following the earlier works in the Matsubara formalism\cite{Ueki2016,Ohuchi2017,Masaki2018charge}. The Green's function in the mixed representation is introduced as the Fourier transform with respect to the relative coordinate $\bm{r} = \bm{r}_{1}-\bm{r}_{2}$ of the phase transformed Green's function $\underline{\widehat{G}}^{M}(\bm{r}_{1},\bm{r}_{2};\ii\omega_{n})=e^{\ii I(\bm{R},\bm{r}_{1})\widehat{\tau}_{3}} \widehat{G}^{\M}(\bm{r}_{1},\bm{r}_{2};\ii \omega_{n})e^{-\ii I(\bm{R},\bm{r}_{2})\widehat{\tau}_{3}}$:
\begin{align}
\widehat{G}^{\M}(\bm{k},\bm{R};\ii \omega_{n})
= \int \dd \bm{r}e^{-\ii \bm{k}\cdot \bm{r}}\underline{\widehat{G}}^{M}(\bm{r}_{1},\bm{r}_{2};\ii\omega_{n}),
\end{align}
where $I(\bm{R},\bm{r}_{1}) = \frac{e}{c}\int_{\bm{r}_{1}}^{\bm{R}}\dd \bm{s}\cdot \bm{A}(\bm{s})$ is defined as a contour integral of the vector potential $\bm{A}$ along the line 
between $\bm{r}_{1}$ and $\bm{R}$, electron charge $e= -|e|$, and the speed of light $c$. 
We consider the gauges where the vector potential is independent of the imaginary time, namely, the electric field is described by the scalar potential.

The quasiclassical Green's function is defined by the energy integral of $\widehat{G}^{\M}(\bm{k}_{\F},\bm{R};\ii \omega_{n})$, where the energy $\xi_{\bm{k}}$ stands for the single particle excitation energy in the normal state and is measured from the Fermi level $\xi_{\F}\equiv \xi_{\bm{k}_{\F}} = 0$.
\begin{align}
\widehat{g}^{\M}(\bm{k}_{\F},\bm{R};\ii \omega_{n})
&=\oint_{C_{\mathrm{qc}}}\dfrac{\dd \xi_{\bm{k}}}{\ii\pi}\widehat{G}^{\M}(\bm{k},\bm{R};\ii\omega_{n})\\
\text{with}~
\widehat{g}^{\M}
&=\begin{bmatrix}
g^{\M} & f^{\M} \\
-\bar{f}^{\M} & \bar{g}^{\M}
\end{bmatrix},
\end{align}
where $C_{\mathrm{qc}}$ stands for the mean of two-contour contributions: one is the counterclockwise contour in the half upper $\xi_{\bm{k}}$-plane, and the other is the clockwise contour in the half lower $\xi_{\bm{k}}$-plane. The transport equation for the quasiclassical Green's function can be obtained through the following steps. (i) Expand the left- and right-sided Gor'kov equations up to the second order in the gradient expansion. (ii) Subtract the right equation from the left equation, which is the transport equation for the Gor'kov Green's function. (iii) Integrate the obtained equation along the above contour while approximating the momentum dependence of the self energy and the pair potential at the Fermi momentum $\bm{k}_{\F}$. We finally obtain the augmented Eilenberger equation
\begin{align}
[\ii\omega_{n}\widehat{\tau}_{3} &+ \widehat{\sigma}^{\M},\widehat{g}^{\M}]_{\star}
+\ii \bm{v}_{\F}\cdot \partial_{\bm{R}}\widehat{g}^{\M} \nonumber \\
&+
\dfrac{e}{2} \left(\bm{v}_{\F} \times \bm{B}\right)\cdot
\left(\ii \dfrac{\partial}{\partial \bm{k}_{\F}}\right)
\left\{\widehat{\tau}_{3}, \widehat{g}^{\M}\right\} = 0, \label{eq-Kita}
\end{align}
where $\bm{v}_{\F} \leftdef \bm{k}_{\F}/m$ is the Fermi velocity. 
The third term including magnetic field $\bm{B} = \Rot \bm{A} = B \hat{e}_{z}$ is called a Hall term, in which $\{\widehat{\tau}_{3},\widehat{g}^{\M}\}$ is the anticommutation of $\widehat{\tau}_{3}$ and $\widehat{g}^{\M}$. Note that $\hat{e}_{\mu}$ stands for the unit vector along the $\mu$-direction, and we take the $z$-direction perpendicular to the 2D system. In the cylindrical coordinate $(R, \theta)$, $\hat{e}_{R} = \bm{R}/R$ and $\hat{e}_{\theta} = \hat{e}_{z} \times \hat{e}_{R}$. 
The self energy matrix in the commutation relation consists of the pair potential $\Delta$ and the impurity self energy $\widehat{\sigma}_{\imp}^{\M}$:
\begin{align}
\widehat{\sigma}^{\M}(\bm{k}_{\F},\bm{R};\ii\omega_{n}) = 
\begin{bmatrix}
\sigma_{\dd}^{\M} & \sigma_{\od}^{\M}\vspace{0.2em} \\
\bar{\sigma}_{\od}^{\M} & \bar{\sigma}_{\dd}^{\M}
\end{bmatrix}&(\bm{k}_{\F},\bm{R};\ii\omega_{n})\nonumber\\
=
\begin{bmatrix}
0 & \Delta(\bm{k}_{\F},\bm{R}) \\
-\Delta^{*}(\bm{k}_{\F},\bm{R})& 0
\end{bmatrix} &+ \widehat{\sigma}_{\imp}^{\M}(\bm{k}_{\F},\bm{R};\ii\omega_{n}).
\end{align}
The details of $\Delta$ and $\widehat{\sigma}_{\imp}^{\M}$ are discussed later.
The symbol used in the first term is defined as
\begin{align}
[A,B]_{\star}
&\equiv
A\star B
-B\star A,
\end{align}
and 
\begin{widetext}
\begin{align}
A\star B 
&\equiv
\lim_{\bm{R}^{\prime} \to \bm{R}}\lim_{\bm{k}_{\F}^{\prime}\to \bm{k}_{\F}
}\exp\left[\dfrac{\ii}{2}\left(\dfrac{\partial}{\partial \bm{k}_{\F}^{\prime}}\cdot \partial_{\bm{R}}
-\dfrac{\partial}{\partial \bm{k}_{\F}}\cdot \partial_{\bm{R}^{\prime}}
\right)\right]
A(\bm{k}_{\F},\bm{R};\ii\omega_{n}) B(\bm{k}_{\F}^{\prime},\bm{R}^{\prime};\ii\omega_{n}).
\end{align}
\end{widetext}
This expression itself includes the higher contribution in the gradient expansion, but we retain only the terms up to the first order. 
The spatial derivative is defined as 
\begin{align}
\partial_{\bm{R}} =
\begin{cases}
\bm{\nabla} &\text{on}~g,\bar{g}, \sigma_{\dd},\bar{\sigma}_{\dd}\\\vspace{0.5em}
\bm{\nabla} -\dfrac{2\ii e}{c}\bm{A}(\bm{R}) &\text{on}~f,\sigma_{\od}\\
\bm{\nabla} +\dfrac{2\ii e}{c}\bm{A}(\bm{R}) &\text{on}~\bar{f}, \bar{\sigma}_{\od}.
\end{cases}
\end{align}

The impurity self energy is treated within the self consistent Born approximation (SCBA), which is given by
\begin{align}
\widehat{\sigma}_{\imp}^{\M}(\bm{k}_{\F},\bm{R};\ii\omega_{n}) = \ii\Gamma_{\n}\braket{\widehat{g}^{\M}(\bm{k}_{\F}^{\prime},\bm{R};\ii\omega_{n})}_{\F^{\prime}} \label{eq-imp}
\end{align}
with the scattering rate in the normal state $\Gamma_{\n}$. The angle bracket on the right-hand side stands for the average on the Fermi surface defined as $\braket{o(\bm{k}_{\F})}_{\F} = (2\pi)^{-1}\int_{0}^{2\pi}\dd \alpha o(\bm{k}_{\F})$. We define $\alpha$ as the angle between $\bm{k}_{\F}$ and $\hat{e}_{x}$. The pair potential is complemented by the gap equation
\begin{align}
\Delta(\bm{k}_{\F},\bm{R}) = g \nu_{\n}(\xi_{\F})\ii \pi T \sum_{n}\Braket{V_{\bm{k}_{\F},\bm{k}_{\F}^{\prime}}^{L_{z}}f^{\M}(\bm{k}_{\F}^{\prime},\bm{R};\ii\omega_{n})}_{\F^{\prime}}. \label{eq-gap}
\end{align}
The type of pairing interactions is characterized by $V_{\bm{k}_{\F},\bm{k}_{\F}^{\prime}}^{L_{z}}$; The superscript specifies the type of pair potentials with a single vortex and corresponds to the total angular momentum of the Cooper pair consisting of chirality and vorticity. We fix the vorticity $+1$, i.e., along the $z$-axis, and label the s-wave pair potential as $L_{z}=1$, and chiral p-wave pair potential with plus and minus chirality as $L_{z} = 2$ and $0$, respectively. The explicit forms of pairing interactions are given by $V_{\bm{k}_{\F},\bm{k}_{\F}^{\prime}}^{L_{z}}= 1$ for $L_{z} = 1$,  and $V_{\bm{k}_{\F},\bm{k}_{\F}^{\prime}}^{L_{z}}= 2\cos (\alpha-\alpha^{\prime})$ for $L_{z} = 2$ or $0$. 
Here $g$ and $\nu_{\n}(\xi_{\F})$ are the coupling constant and the density of states in the normal state at the Fermi level, which are related to transition temperature $T_{\mathrm{c}}$ as
\begin{align}
\dfrac{1}{g\nu_{\n}(\xi_{\F})} = \ln\dfrac{T}{T_{\mathrm{c}}} + 2\pi \sum_{0\le n\le N_{\mathrm{c}}}\dfrac{1}{2n+1} 
\end{align}
with cut-off $N_{\mathrm{c}}$. The integer $N_{\mathrm{c}}$ is defined as the largest integer $n$ satisfying $\omega_{n}\le \omega_{\mathrm{c}}$ and we set $\omega_{\mathrm{c}}$ to $40T_{\mathrm{c}}$ in this paper. For a chiral p-wave SC, there are induced components in addition to the dominant component and the pair potential with an axisymmetric vortex is described by 
\begin{align}
\Delta(\bm{k}_{\F},\bm{R})  = \Delta_{+}(R) e ^{\ii (\alpha - \theta) + \ii L_{z}\theta} + \Delta_{-}(R) e ^{-\ii (\alpha - \theta) + \ii L_{z}\theta}.
\end{align}
Here $\Delta_{+}(R\to \infty) \neq 0$ and  $\Delta_{-}(R\to \infty) = 0$ for $L_{z} = 2$, 
while $\Delta_{+}(R\to \infty) = 0$ and  $\Delta_{-}(R\to \infty) \neq 0$ for $L_{z} = 0$. We call the component which is zero for $R\to \infty$ the induced component.

We write down the augmented Eilenberger equation order by order in quasiclassical parameter $1/(k_{\F}\xi_{0})$, the small parameter of the gradient expansion. We introduce a coherence length $\xi_{0}\equiv v_{\F}/\Delta_{0}$, where $v_{\F} =|\bm{v}_{\F}|$ and  $\Delta_{0}$ is the energy gap in the bulk at $T=0$. The self-energy form \eqref{eq-imp} and gap equation \eqref{eq-gap} hold the same form order by order. We expand a quantity $\mathcal{O}$ as $\mathcal{O}_{0} + \mathcal{O}_{1} + \cdots $, and $\mathcal{O}$ can be any of $\widehat{g}$,  $\widehat{\sigma},$\footnote{
Here the order of the subscripts of $\sigma$ can be exchanged, namely, perturbative expansion of matrix is equivalent to the perturbative expansion of each component. Thus we write components of self energy terms as $\sigma_{\od,0}$, $\sigma_{\imp,\od,0}$ rather than 
$\sigma_{0,\od}$, $\sigma_{\imp,0,\od}$
} and $\Delta$. The augmented Eilenberger equation of the zeroth order is reduced to the conventional Eilenberger equation:
\begin{align}
2\ii \omega_{n} f_{0}^{\M} + \ii \bm{v}_{\F} \cdot \partial_{\bm{R}} f_{0}^{\M} +2( \sigma_{\dd,0}^{\M} f_{0}^{\M} - \sigma_{\od,0}^{\M} g_{0}^{\M})= 0.
\end{align}
We solve this equation using Riccati transformation and the normalization condition is given by
$g_{0}^{\M} = \mathrm{sgn}(\omega_{n} )\left[1- f_{0}^{\M}\bar{f}_{0}^{\M}\right]^{1/2}$.\cite{Nagato1993,Schopohl1995,Schopohl1998}
The first order equation consists of two closed parts: One is an equation for $g_{1}^{\M} +\bar{g}_{1}^{\M}$, and the other is a coupled equation for $g_{1}^{\M} - \bar{g}_{1}^{\M}$, $f_{1}^{\M}$, and $\bar{f}_{1}^{\M}$.
The charging effect, as we see next, is described by $g_{1}^{\M} + \bar{g}_{1}^{\M}$, which obeys
\begin{widetext}
\begin{align}
\dfrac{1}{2}\ii v_{\F} \dfrac{\partial (g_{1}^{\M} + \bar{g}_{1}^{\M})}{\partial s} 
=& \ii \dfrac{ev_{\F}B}{k_{\F}} \dfrac{\partial g_{0}^{\M}}{\partial \alpha} + 
\dfrac{\ii \hat{e}_{b}}{2k_{\F}}\cdot\left[2\left(\partial_{\bm{R}}g_{0}^{\M} \dfrac{\partial \sigma_{\dd,0}^{\M}}{\partial \alpha}-\partial_{\bm{R}}\sigma_{\dd,0}^{\M} \dfrac{\partial g_{0}^{\M}}{\partial \alpha}
\right)\right.
\nonumber \\
&\hspace{3.5em}+\left.\partial_{\bm{R}}\sigma_{\od,0}^{\M} \dfrac{\partial \bar{f}_{0}^{\M}}{\partial \alpha}
-\partial_{\bm{R}}\bar{\sigma}_{\od,0}^{\M} \dfrac{\partial f_{0}^{\M}}{\partial \alpha}
-\dfrac{\partial \sigma_{\od,0}^{\M}}{\partial \alpha} \partial_{\bm{R}}\bar{f}_{0}^{\M}
+\dfrac{\partial \bar{\sigma}_{\od,0}^{\M} }{\partial \alpha}\partial_{\bm{R}}f_{0}^{\M}
\right]. \label{eq-g1}
\end{align}
\end{widetext}
Here $\hat{e}_{b} \leftdef \hat{e}_{z} \times \hat{e}_{s}$ with $\hat{e}_{s} \leftdef \bm{v}_{\F} /v_{\F}$, and $s$ is the coordinate along $\hat{e}_{s}$. In the following, we neglect the Hall term, which is valid when the system is in the type-II limit.
 We see that there is a solution satisfying the boundary condition that $g_{1}(R\to \infty) = \bar{g}_{1}(R\to\infty) = 0$, i.e., the integration of Eq.~\eqref{eq-g1} over $s$ from $-\infty$ to $\infty$ is zero. We obtain $(g_{1}+\bar{g}_{1})/2$ by integrating Eq.~\eqref{eq-g1} from a point at infinity along the trajectory specified by $\alpha$ and an impact parameter $b$, the coordinate along $\hat{e}_{b}$. 

Next we see the form of charge density $\rho(\bm{R})$ in 2D given by\cite{Eliashberg1972, Serene1983}
\begin{align}
\rho(\bm{R})
& \leftdef e (n(\bm{R})-n_{\n})\nonumber\\
&= -e \nu_{\n}(\xi_{\F})\ii \pi T \sum_{n} \Tr\left[\Braket{\widehat{g}^{\M}}_{\F} - \Braket{\widehat{g}_{\n}^{\M}}_{\F}\right]\nonumber\\
& -2\nu_{\n}(\xi_{\F}) e^{2}\Phi(\bm{R}). \label{eq-rho}
\end{align}
Here $g_{\n}^{\M}$ and $n_{\n}$ describe, respectively, the quasiclassical Green's function and particle density in normal-state without electromagnetic fields, and the Matsubara sum of $g_{\n}^{\M}$ vanishes.  The scalar potential $\Phi$ in the last term, whose gradient leads to the electric field, is a high-energy contribution.  In other words, the origin of the scalar potential is in the limiting procedure that chemical potential $\mu$ is taken to infinity, which is accordingly equivalent to the exchange of the order of the energy integral and the Matsubara frequency sum, as seen explicitly in the following:
In the mixed representation with the gradient expansion, the Gor'kov Green's function in the normal state but in the presence of the scalar potential is given by 
$G_{\n,\Phi}^{\M}(\bm{k},\bm{R};\omega_{n}) = (\xi_{\bm{k}} + e\Phi(\bm{R}) - \ii \omega_{n})^{-1}$. 
The momentum integration can be reduced to 
$\int \dd \bm{k}/(2\pi)^{2} o(\bm{k}) = \int_{-\mu\to-\infty}^{\infty} \dd \xi \nu_{\n}(\xi)\braket{o(\bm{k})}_{\xi}$ 
with $\braket{o(\bm{k})}_{\xi} =( \nu_{\n}(\xi))^{-1}\int \dd \bm{k}/(2\pi)^{2}o(\bm{k})\delta(\xi-\xi_{\bm{k}})$. 
Note  the lower limit of the integral. In this case the momentum average is trivial, and we approximate the density of state as the value at the Fermi level. 
We can calculate the following quantity:
\begin{widetext}
\begin{align}
&\left[\int_{-\infty}^{\infty} \dfrac{\dd \xi_{\bm{k}}}{\ii \pi}(\ii \pi T)\sum_{n}- (\ii \pi T)\sum_{n}\int_{-\infty}^{\infty} \dfrac{\dd \xi_{\bm{k}}}{\ii \pi}\right]\left(G_{\n,\Phi}^{\M}(\bm{k},\bm{R};\ii \omega_n)-G_{\n}^{\M}(\bm{k},\bm{R};\ii \omega_n)\right)\nonumber \\
&\simeq \int_{-\infty}^{\infty} \dfrac{\dd \xi_{\bm{k}}}{2}
\left[\tanh\left(\dfrac{\xi_{\bm{k}} + e \Phi(\bm{R})}{2T}\right)-\tanh\left(\dfrac{\xi_{\bm{k}}}{2T}\right)\right]\simeq e\Phi(\bm{R}).
\end{align}
\end{widetext}

Equation~\eqref{eq-rho} includes the scalar potential which should be determined through the complementary Poisson equation given by $\nabla^{2} \Phi(\bm{R}) = -\frac{4\pi \rho(\bm{R})}{d}$.\cite{Artemenko1979}
The combination of this and Eq.~\eqref{eq-rho} gives us the inhomogeneous differential equation for $\Phi$ as follows:
\begin{align}
\left[\nabla^{2} - \lambda_{\TF}^{-2}\right]\Phi(\bm{R}) =  -\dfrac{4\pi}{d}\rho_{0}(\bm{R}),\\
\rho_{0}(R) \equiv  \dfrac{d T}{4 \lambda_{\TF}^{2} e} \sum_{\omega_{n} > 0 }\Braket{\mathrm{Im}  \left[\Tr \widehat{g}_{1}^{\M}
(\bm{k}_{\F},\bm{R}; \ii \omega_{n})\right]}_{\F},
\end{align}
where $\lambda_{\TF}\equiv  (d/(8\pi \nu_{\n}(\xi_{\F}) e^{2}))^{1/2}$ is the Thomas--Fermi length. Note that $\rho_{0}$ is linear order in $1/(k_{\F}\xi_{0})$.
\begin{figure*}[t]
\centering
\includegraphics[width = 50 em]{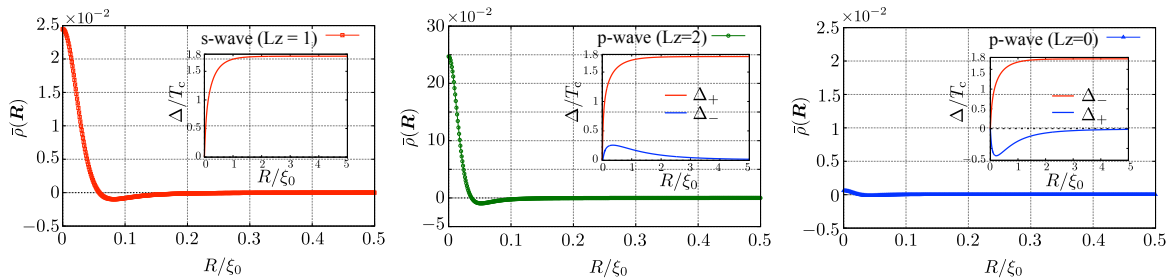}
\caption{Charge density profiles for (a) the s-wave SC and the chiral p-wave SCs with (b) the parallel vortex and (c) the antiparallel vortex. The inset of each panel shows the radial profile of the pair potential.}
\label{fig-charging}
\end{figure*}
\section{Numerical Results}\label{numerics}
In this section we show the numerical results of charge densities in vortex cores.
We normalize the 2D charge density by $2|e|\nu_{\n}(\xi_{\F})T_{\mathrm{c}}$, and introduce the notation for the normalized charge density as $\bar{\rho} = \rho/(2|e|\nu_{\n}(\xi_{\F})T_{\mathrm{c}})$. In the first subsection, we show the typical properties of vortex charge-densities and that there are charged and uncharged vortices.

\subsection{charged and uncharged vortices}
First of all, we investigate typical behaviors of charging effects.
Figures~\ref{fig-charging}(a), (b), and (c), respectively, show that the profiles of charge densities of vortices for
s-wave SC ($L_{z} = 1$) and chiral p-wave SCs ($L_{z} = 2$ and $0$ ). Temperature is set to $0.1 T_{\mathrm{c}}$ and 
impurity is absent. The s-wave SC and the chiral p-wave SC of $L_{z} = 2$ have charged vortices, while that of $L_{z}=0$ 
has an uncharged vortex. 

The spatial derivative has two contributions:  $\hat{e}_{b}\cdot \hat{e}_{R}\partial/\partial R$ and $\hat{e}_{b} \cdot \hat{e}_{\theta} R^{-1}\partial/\partial_{\theta}$. We call the former {\it radial part} and the latter {\it angular part}. We can show through an explicit calculation that the angular part is proportional to the total angular momentum $L_{z}$. Thereby the angular part vanishes when $L_{z} = 0$. This is equivalent to the cancelation of phases from chirality and vorticity. 

The radial part includes the impact parameter $b$ in its numerator, which stands for the angular momentum of the quasiparticle, and the contribution from the core center becomes small. 
The contribution from the angular part for non-zero $L_{z}$ is dominant over that from the radial part since the radial part vanishes when the quasiclassical trajectory passes through the origin $b=0$; Note that the charge density at the origin of $L_{z}= 0$ is due to the screening effect. Thus the vortex is charged for nonzero $L_{z}$ and (almost) uncharged for $L_{z}=0$. 

In addition, we discuss the magnitude of the charge densities for the two charged vortices. We see from Fig.~\ref{fig-charging} that the charge density of the vortex with $L_{z} = 2 $ is one order of magnitude, rather than twice, greater than that  with $L_{z}=1$. If we neglect the induced component of the pair potential for $L_{z}=2$, we see that the charge of the vortex with $L_{z}=2$ is almost twice as much as that with $L_{z}=1$ in Ref.~\onlinecite{Masaki2018charge}. The induced component drastically changes the quasiclassical Green's function, which enhances the charge density. This holds also for $T= 0.2T_{\mathrm{c}}$.

\begin{figure}
\includegraphics[width = 20em]{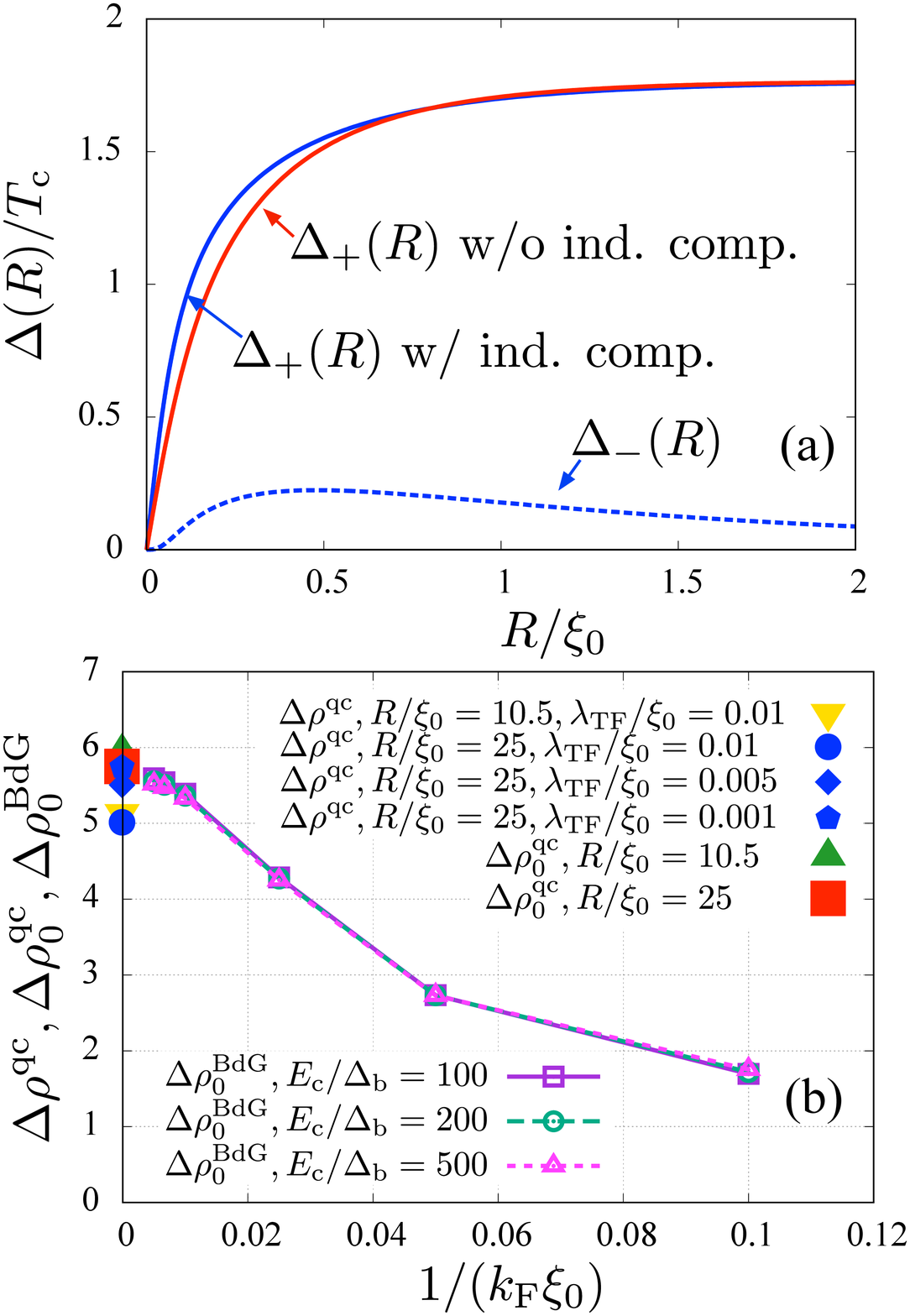}
\caption{(a) The profiles of pair potentials. The red line is $\Delta_{+}$ in the absence of $\Delta_{-}$, while the blue solid line is $\Delta_{+}$ in the presence of $\Delta_{-}$, which is shown by the blue dashed line. (b) $[1/(k_{\F}\xi_{0})]$-dependence of the charge-density-ratios defined in the text. 
$R$-dependence is seen from $\Delta\rho_{0}^{\mathrm{qc}}$, and $\Delta\rho^{\mathrm{qc}}$ for $\lambda_{\TF}/\xi_{0} =0.01$. The sequence of the blue symbols represents that their limiting value is $\Delta \rho_{0}^{\mathrm{qc}}$ displayed by the red square.}
\label{fig-BdGvsQC}
\end{figure}

To confirm this validity, we study the enhancement caused by the induced component of the pair potential also using microscopic BdG equation for $T= 0.2T_{\mathrm{c}}$ in the left part of this subsection. We first review the scheme of BdG equation\cite{MatsumotoHeeb2001,Machida2002}. We use the following approximations since the direct numerical calculation has considerable computational cost:
(A) We apply the profiles of pair potential obtained within the quasiclassical theory to those in the BdG equation and do not perform the self-consistent calculation using the gap equation. (B) We neglect the magnetic field $\bm{B} = 0$, which is the same situation as in the present quasiclassical theory. (C) We consider the linear response for the charge density within the local (Thomas--Fermi) approximation. Using the local approximation we write down the charge density as 
\begin{align}
e (n(\bm{R})) 
&= 2e\sum_{\nu} |u_{\nu}^{0}(\bm{R})|^{2}f(E_{\nu}^{0}) + \int \dd \bm{R}^{\prime} \chi(\bm{R},\bm{R}^{\prime})\Phi(\bm{R}^{\prime})\nonumber\\
&\sim
en_{0}(\bm{R}) - \dfrac{d}{4\pi} \lambda_{\TF}^{-2}\Phi(\bm{R}),\label{eq-charge-BdG}
\end{align}
where $\chi(\bm{R},\bm{R}^{\prime})$ is the density-density correlation function\cite{Machida2002}. We use the eigenfunctions $u_{\nu}^{0}(\bm{R})$ and the eigenenergies $E_{\nu}^{0}$ of the BdG equation in the absence of the scalar potential given by
\begin{align}
\widehat{H}_{\mathrm{BdG}}(\bm{R})
\begin{pmatrix}
u_{\nu}^{0}(\bm{R}) \\
v_{\nu}^{0}(\bm{R})
\end{pmatrix}
=E_{\nu}^{0}
\begin{pmatrix}
u_{\nu}^{0}(\bm{R}) \\
v_{\nu}^{0}(\bm{R})
\end{pmatrix}
\end{align}
with $\int \dd \bm{R}[u_{\nu^{\prime}}^{0*}(\bm{R})u_{\nu}^{0}(\bm{R})+ v_{\nu^{\prime}}^{0*}(\bm{R})v_{\nu}^{0}(\bm{R})] = \delta_{\nu^{\prime}\nu}$,
\begin{align}
\widehat{H}_{\mathrm{BdG}}(\bm{R})
&=
\begin{pmatrix}
H_{0}(\bm{R}) &
D(\bm{R}) \\
-[D(\bm{R})]^{*} &
 - [H_{0}(\bm{R})]^{*}
\end{pmatrix},\\
h_{0}(\bm{R}) &= \dfrac{-\nabla^{2}-k_{\F}^{2}}{2m},\\
D(\bm{R})&=  \sum_{s=\pm} \dfrac{1}{\ii k_{\F}}\left[
\dfrac{1}{2}\dfrac{\partial \Delta_{s}(\bm{R})}{\partial R_{s}} 
+\Delta_{s}(\bm{R})\dfrac{\partial }{\partial R_{s}} \right],\label{eq-delta-sum}\\
\Delta_{\pm}(\bm{R}) &= \Delta_{\pm}(R) e ^{(2\mp 1) \ii\theta},\\
\dfrac{\partial}{\partial R_{\pm}} &= (\hat{e}_{x} \pm \ii \hat{e}_{y})\cdot \dfrac{\partial }{\partial \bm{R}}.
\end{align}
Here the spatial profiles $\Delta_{\pm}(R)$ are obtained in the quasiclassical scheme. When we neglect the induced component of the pair potential, we consider only the contribution $s = +$ in Eq.~\eqref{eq-delta-sum}. Note that $\Delta_{+}(R)$ exhibits difference between profiles in the cases with and without the induced component $\Delta_{-}(R)$, as shown in Fig.~\ref{fig-BdGvsQC}(a). 
The BdG Hamiltonian is block-diagonalized with respect to the angular momentum of quasiparticles  $l$. We use the Fourier--Bessel expansion to obtain the eigenenergies and eigenstates from the differential equations along the radial direction. An eigenfunction takes the form $(u_{\nu}^{0}(\bm{R}),v_{\nu}^{0}(\bm{R}))=(u_{l,n}^{0}(R)e^{\ii (l+1)\theta},v_{l,n}^{0}(R)e^{\ii (l-1)\theta})$ and $\nu$ is specified by $l$ and radial index $n$.
Concerning the first term of Eq.~\eqref{eq-charge-BdG}, the summation index runs over  eigenstates with positive and negative eigenenergies. We introduce an energy cut-off $E_{\mathrm{c}}$ which restricts the summation range as $|E_{\nu}| < E_{\mathrm{c}} $.  In Eq.~\eqref{eq-charge-BdG}, we have approximated $\chi(\bm{R},\bm{R}^{\prime})\sim -\frac{d}{4\pi}\lambda_{\TF}^{-2}\delta(\bm{R}-\bm{R}^{\prime})$ to obtain the last form. The scalar potential in Eq.~\eqref{eq-charge-BdG} is determined using the complementary Poisson equation as in the quasiclassical theory:
\begin{align}
(\nabla^{2} - \lambda_{\TF}^{-2} )\Phi(\bm{R}) &=-\dfrac{4\pi}{d} e(n_{0}(\bm{R}) - n_{\infty})\nonumber\\
&\equiv -\dfrac{4\pi}{d} \rho_{0}(\bm{R})\\
\mathrm{with}~n_{\infty} &\equiv \lim_{R\to \infty} n_{0}(\bm{R}).
\end{align}
These quantities are independent of the direction and we omit $\theta$ from their arguments in the following.
The Poisson equation requires two boundary conditions. They are usually given by Dirichlet-type conditions at two boundaries, and thus it is desirable to have the source term for sufficiently large range of the radial coordinate. However there is enormous numerical cost to calculate the charge density $n_{0}(R)$ since contributions from eigenstates with high angular momentum becomes larger in the further region from the vortex core. Thus we introduce an additional approximation\cite{Artemenko1979}, in which we solve the Poisson equation and obtain the charged density as
\begin{align}
\Phi(R) &\sim -\dfrac{4\pi  \lambda_{\TF}^{2}}{d} \rho_{0}(R),\\
\rho(R) &= -\dfrac{d}{4\pi}\nabla^{2}\Phi(R) \sim   \lambda_{\TF}^{2}\nabla^{2}\rho_{0}(R).
\end{align}
The approximation is justified when $\lambda_{\TF}$ is sufficiently small compared with the characteristic length scale of the charge density, and it is the slope of the dominant component of the pair potential $\xi_{1}\equiv \lim_{R\to 0} \Delta_{+}(R)/R$. 
Particularly we focus on the charge density at the origin, which allows us to evaluate it as $\rho (R=0) \sim 2\lambda_{\TF}^{2} \partial^{2}\rho_{0}(R=0)/\partial R^{2}$. 
In the BdG scheme, we calculate the charge density only in this way, i.e., it is in the limit $\lambda_{\TF}\to 0$. 
On the other hand, we consider both $\rho(R)$ itself and $\lambda_{\TF}^{2}\nabla^{2}\rho_{0}(R)$ in the quasiclassical theory to study the effects of finite $\lambda_{\TF}$. 

In Fig.~\ref{fig-BdGvsQC}(b), we show as a function of the quasiclassical parameter $1/(k_{\F}\xi_{0})$ the ratios of the charge density at the origin with the induced component $\Delta_{-}$ to that without the absence of $\Delta_{-}$, which is described as $\Delta \rho \equiv \rho_{\mathrm{w/}}(R=0)/\rho_{\mathrm{w/o}}(R=0)$. We distinguish the way of calculation of $\Delta \rho$ by using subscript and superscript of $\Delta \rho$.
The subscript takes blank or 0 corresponding to whether we use $\rho$ or $\lambda_{\TF}^{2}\nabla^{2}\rho_{0}(R)$, while the superscript takes qc or BdG corresponding to the framework. 
There are several remarks: (1) We do not calculate $\Delta \rho_{}^{\BdG}$ in this paper as mentioned above. (2) $\Delta \rho^{\mathrm{qc}}$ and $\Delta \rho_{0}^{\mathrm{qc}}$ are independent of $k_{\F}\xi_{0}$ when $\bm{B}=0$, because we calculate the charge density up to the first order in $1/(k_{\F}\xi_{0})$ within the quasiclassical theory, and thus we display it at $1/(k_{\F}\xi_{0}) = 0$. (3) $\Delta \rho_{0}^{\mathrm{qc}/\mathrm{BdG}}$ is independent of $\lambda_{\mathrm{TF}}$ and we corroborate $\lim_{\lambda_{\mathrm{TF}\to 0}}\Delta \rho^{\mathrm{qc}} = \Delta \rho_{0}^{\mathrm{qc}}$ in Fig.~\ref{fig-BdGvsQC}(b). 
We then find $\lim_{k_{\F}\xi_{0}\to \infty}\Delta \rho_{0}^{\BdG} = \Delta \rho_{0}^{\mathrm{qc}}$ in Fig.~\ref{fig-BdGvsQC}(b) and confirm our finding about the enhancement of charge density for $L_{z} = 2 $ compared with that for $L_{z} = 1$. We will comment on the validity of the quasiclassical approach in the next subsection. The $k_{\F}\xi_{0}$-dependence of $\Delta \rho_{0}^{\BdG}$ stems from higher order contributions in $1/(k_{\F}\xi_{0})$ to the charge densities, and it cannot be described by the augmented quasiclassical theory.

\subsection{temperature dependence}
\begin{figure}[t]
\centering
\includegraphics[width = 25 em]{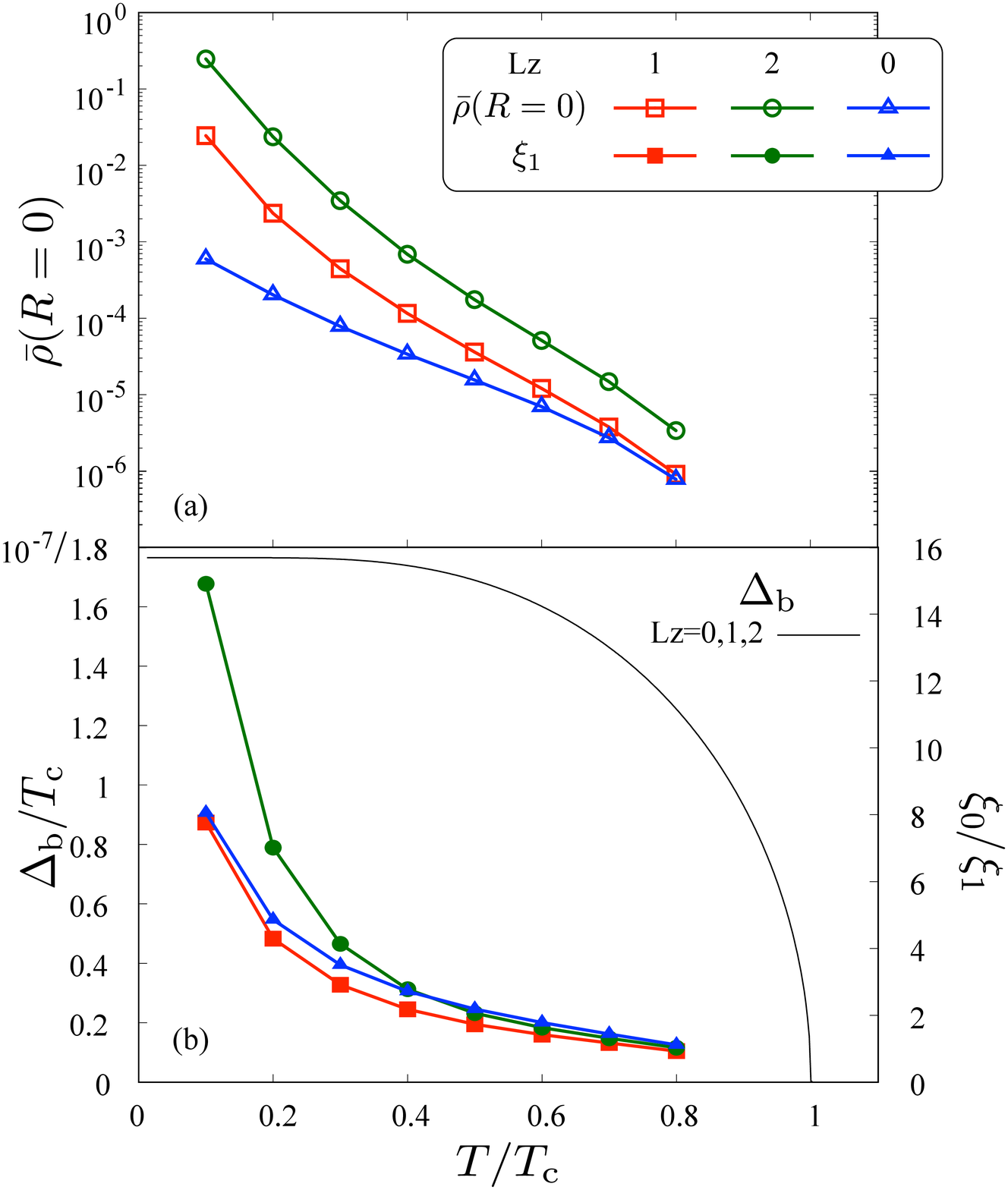}
\caption{(a) Temperature dependence of the charge density at the origin. (b) Temperature dependence of the energy gap in the bulk and the coherence length as the slope of the pair potential at the origin, $\xi_{1}$.}
\label{T-dep}\end{figure}
Next we discuss the temperature dependence of charge densities. In the s-wave case, the temperature dependence of the charge density at the origin is studied in Ref.~\onlinecite{Ohuchi2017}. In a similar way, we study the dependence of chiral p-wave cases.
Figure~\ref{T-dep}(a) shows the temperature dependence of the charge density at the origin. The decrease in low temperature is quite drastic for $L_{z} = 1$ and $2$, which suggests that the low energy bound states are essential, as studied in the earlier work using the BdG equation\cite{MatsumotoHeeb2001}. Such kind of contribution is absent for $L_{z}=0$ and thus the difference between charged and uncharged vortices is present for lower temperature. The reductions for the higher temperature are similar for all the cases and we speculate that the charge densities for high temperatures are due to the spatial modulation of extended states and the screening effects via the Poisson equation. 

Note that temperature dependence in low-temperature side is different from that obtained by the BdG equation in the case of the s-wave SC\cite{Hayashi1998}. 
In Ref.~\onlinecite{Hayashi1998}, the charge density increases with decreasing temperature and suddenly saturates at the temperature around the minigap, which is the level spacing of the bound states localized in the vortex core. This saturation cannot be obtained by the quasiclassical theory up to the first order in $1/(k_{\F}\xi_{0})$. Our results of $\rho/(2e\nu_{\n}(\xi_{\F})T_{\mathrm{c}})$ linearly depend on $1/(k_{\F}\xi_{0})$ for the whole region of the temperature. On the other hand, the charge density by the BdG equation has two regimes: $1/(k_{\F}\xi_{0})$-dependence of $\rho_{0}(R=0)/n_{\infty}$, where $n_{\infty}\simeq n_{\n}$, is linear at low temperatures and quadratic at high temperatures. By noting that $2e\nu_{\n}(\xi_{\F})T_{\mathrm{c}}/n_{\n}= O(1/(k_{\F}\xi_{0}))$, we see that our results on temperature dependence is consistent with the charge density obtained using the BdG equation at high temperature compared with the minigap. Even though the scalar potential and the charge density are not self-consistently determined, this physical picture will not be changed. 

The Kramer-Pesch effect is  important because the charge density strongly depends on the slope of the pair potential around the vortex core. In the pure system, the vortex core radius $\xi_{1}$ goes to zero linearly as the temperature does in the quasiclassical theory, while the radius suddenly saturates in the low temperature region in the framework of the BdG equation\cite{kramer1974z,Volovik1993,Hayashi1998a} (We should remark that the gap profiles are calculated using the conventional quasiclassical theory in this paper). The saturated value of $\xi_{1}$ is about $1/k_{\F}$, reflecting the minigap structure.  
Thus at low temperatures compared with the minigap, the Kramer--Pesch effect is inconsistent between the quasiclassical theory and the BdG theory in the pure system. We discuss the Kramer--Pesch effects  later in the presence of impurities. 

The inconsistency in the Kramer--Pesch shrinking can be related  to the inconsistency in the $(1/(k_{\F}\xi_{0}))$-dependence of the charge density at low temperatures. The energy gap in the bulk region is almost saturated in the low temperature region and its value hardly depends on the $1/(k_{\F}\xi_{0})$, while $k_{\F}\xi_{0}$-dependence appears in the core region. The core radius which  we roughly estimate as $\xi_{1}/\xi_{0}\sim1/(k_{\F}\xi_{0})$ contributes to the gradient term, and an effective quasiclassical parameter is of order $O(1)$. The charge density $\rho/(2e\nu_{\n}(\xi_{\F})T_{\mathrm{c}})$ is independent of $1/(k_{\F}\xi_{0})$ when the Kramer--Pesch shrinking is saturated as $\xi_{1}\sim1/k_{\F}$, namely $\rho/n_{\n} \sim O(1/(k_{\F}\xi_{0}))$. Since the saturation of the Kramer--Pesch shrinking is not incorporated in our perturbative approach, the augmented quasiclassical theory cannot describe the saturation of the charge density at low temperatures.

We briefly comment on the validity of calculation in the end of the previous subsection. The temperature is $0.1T_{\mathrm{c}}$, while the $1/(k_{\F}\xi_{0})$ is used down to 0.005 in the BdG equation. For such $1/(k_{\F}\xi_{0})$, we can estimate $\xi_{1}/\xi_{0} \sim T/T_{\mathrm{c}}$ and the practical parameter $1/(k_{\F}\xi_{1}) \sim 0.05$. Therefore our calculation using the BdG equation gives values close to the value obtained by the quasiclassical theory.  

\subsection{impurity effects}
\begin{figure*}[t]
\centering
\includegraphics[width = 55 em]{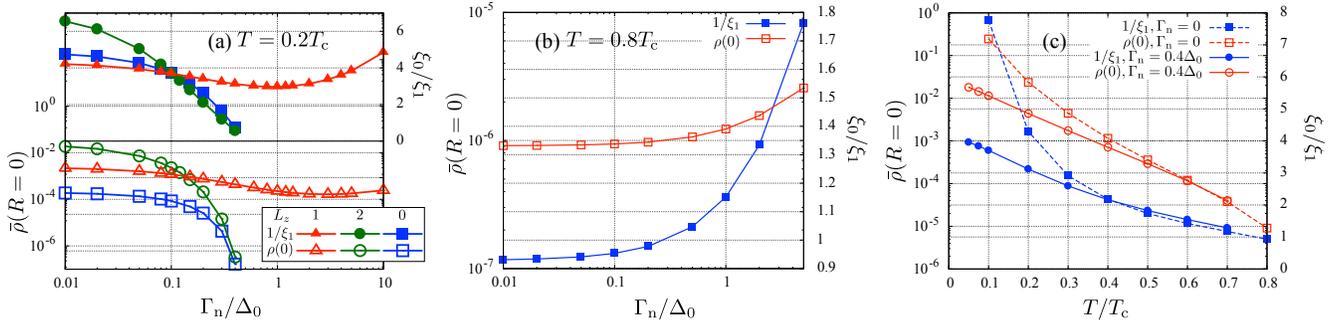}
\caption{
(a) $\Gamma_{\n}$ dependence of the charge density at the origin, and corresponding changes of the initial slopes of the pair potential at the origin $T = 0.2 T_{\mathrm{c}}$. 
(b) $\Gamma_{\n}$ dependence of the charge density and the initial slopes for the pair potential of the s-wave SC at the origin for $T = 0.8 T_{\mathrm{c}}$.
(c) Temperature dependence of the charge density and the slope at the origin for s-wave SCs with and without impurities. }
\label{fig:imp-dep}
\end{figure*}

In this subsection, we study the impurity effects using the quasiclassical theory.
The large charge density results from the large energy gap in the bulk and the sharp slopes of the pair potential around the core.
The scattering of the bound states in the core leads to a gentle slope, while the scattering in the bulk destroys the phase coherence and reduces the energy gap, which correspondingly leads to the longer healing length of the pair potential. In the case of $L_{z} = 2$ has both effects and its vortex charge is drastically reduced. For $L_{z}=0$, scattering effects on bulk quasiparticles appear for relatively large $\Gamma_{\n}/\Delta_{0}$.

In the case of $L_{z} = 1$, the charge density is reduced because of the suppression of the Kramer--Pesch effect by the scattering of the core states for weak $\Gamma_{\n}$, which is up to $\Gamma_{\n} \lesssim \Delta_{0}$. The energy gap in the bulk is robust against the impurity and the decrease of the charge density is different from $L_{z} = 2$. 
For strong $\Gamma_{\n}$, the mean free path becomes shorter than the coherence length and the practical coherence length analogous to the Pippard coherence length becomes shorter as well\cite{Hayashi2005,Hayashi2013b}. This results in the sharp slope of the pair potential at the core, and leads to the enhancement of the vortex charge.

The recovery of the charge density can be seen also at $T=0.8T_{\mathrm{c}}$ in Fig.~\ref{fig:imp-dep}(b). In this case the initial slope monotonically increases as an impurity concentration of Born-type scatters and correspondingly the charge density increases, although these enhancement are less than one order of magnitude.

In Fig.~\ref{fig:imp-dep}(c), we show how impurities affect the temperature dependence of the charge density. Here we set $1/(k_{\F}\xi_{0}) = 0.1$ and $\lambda_{\TF}/\xi_{0} = 0.01$.
For high temperature $T\gtrsim 0.5T_{\mathrm{c}}$, impurity does not affect very much. This is because the charge density and the coherence length are dominated by the bulk quasiparticles, and they are robust because of Anderson's theorem. On the other hand, enhancement of the charge density and the initial slope are suppressed by impurity scattering at low temperatures as a result of the suppression of the Kramer--Pesch shrinking\cite{Hayashi2005}. In the low temperature, the initial slope is dominated by the Andreev bound states, which are affected by the Born-type scatterers. When $\Gamma_{\n} > \Delta/(k_{\F}\xi_{0})$, the Kramer--Pesch shrinking are not determined by the discreteness of the spectra but by impurity scattering, and thus the quasiclassical approximation can be valid. This is also essential to discuss the local density of states in the next section.

\subsection{Local density of states}

We finally investigate the local density of states. The retarded Green's function suffices for this purpose but it is fully phase transformed as
$\underline{\widehat{G}}^{\R}(\vec{x}_{1},\vec{x}_{2})\equiv e^{\ii I(\vec{X},\vec{x}_{1})\widehat{\tau}_{3}} \widehat{G}^{\R}(\vec{x}_{1},\vec{x}_{2})e^{-\ii I(\vec{X},\vec{x}_{2})\widehat{\tau}_{3}}$
 for gauge invariant density of states. Its Wigner representation is given by
\begin{align}
\widehat{G}^{\R}(\bm{k},\bm{R};\omega,T)
= \int \dd \vec{x}e^{\ii\omega t-\ii \bm{k}\cdot \bm{r}}\underline{\widehat{G}}^{\R}(\vec{x}_{1},\vec{x}_{2}), \label{eq-phase-real}
\end{align}
where $\vec{x} = (\bm{r},ct) =\vec{x}_{1}-\vec{x}_{2}$ and $\vec{X} = (\bm{R},cT) = (\vec{x}_{1}+\vec{x}_{2})/2$. The phase factor is defined by $I(\vec{X},\vec{x}_{1}) = \frac{e}{c}\int_{\vec{x}_{1}}^{\vec{X}}\dd \vec{s}\cdot \vec{A}(\vec{s})$ with $\vec{A}(\vec{s}) = (-\bm{A}(\vec{s}),\Phi(\vec{s}))$. The quasiclassical Green's function is defined by the $\xi_{\bm{k}}$-integration of $\widehat{G}^{\R}(\bm{k},\bm{R};\omega,T)$:
\begin{align}
\widehat{g}^{\R}(\bm{k}_{\F},\bm{R};\omega,T) = \oint_{C_{\mathrm{qc}}} \dfrac{\dd \xi_{\bm{k}}}{\ii\pi}\widehat{G}^{\R}(\bm{k}_{\F},\bm{R};\omega,T).
\end{align}
The augmented Eilenberger equation is derived as follows:
\begin{align}
\left[\omega\widehat{\tau}_{3} + 
\widehat{\sigma}^{\R},
\widehat{g}^{\R}\right]_{\star}
&+\ii \bm{v}_{\F}\cdot \partial_{\bm{R}}
\widehat{g}^{\R}+
\dfrac{e}{2}\left[ \left(\bm{v}_{\F}\cdot \bm{E}\right)
\left(-\ii \dfrac{\partial}{\partial\omega}\right)\right.
\nonumber \\
&\hspace{-2em}+\left.
 \left(\bm{v}_{\F} \times \bm{B}\right)
\left(\ii \dfrac{\partial}{\partial \bm{k}_{\F}}\right)\right]
\left\{\widehat{\tau}_{3}, \widehat{g}^{\R}\right\} = 0, \label{eq-Kita-realf}
\end{align}
The Moyal product in the first term does not have to be changed since we consider the equilibrium case and the time derivative does not affect.
Note that the third term appears for the phase factor in Eq.~\eqref{eq-phase-real}, which is absent in the Matsubara formalism. We use the perturbative way to solve Eq.~\eqref{eq-Kita-realf},\cite{Levanda1994,PhysRevB.64.054503} where we  obtain $g_{1}^{\R}$ using the assumption that $g_{1}^{\R} = \bar{g}_{1}^{\R}$. The difference between them does not contribute to the asymmetry of the local density of states if it is finite. The electronic local density of states are defined by
\begin{align}
\nu_{\mathrm{e}}(\bm{R},\omega) 
&= \nu_{\n}(\xi_{\F})\braket{(g^{\R}-g^{\A})(\bm{k}_{\F},\bm{R};\omega)}_{\F}\nonumber\\
&=2\nu_{\n}(\xi_{\F}) \mathrm{Re}\braket{(g^{\R})(\bm{k}_{\F},\bm{R};\omega)}_{\F}
\end{align}
By defining the hole local density of states as 
$\nu_{\mathrm{h}}(\bm{R},\omega) = -\nu_{\n}(\xi_{\F})\braket{(\bar{g}^{\R}-\bar{g}^{\A})(\bm{k}_{\F},\bm{R};\omega)}_{\F} $, we obtain the charge density as 
\begin{align}
\rho(\bm{R}) 
&= \int \dfrac{\dd \omega}{2} \left[\nu_{\mathrm{e}}(\bm{R},\omega) \tanh\dfrac{\omega + e\Phi(\bm{R})}{2T} \right.\nonumber \\ &\hspace{5em}\left.-\nu_{\mathrm{h}}(\bm{R},\omega) \tanh\dfrac{\omega - e\Phi(\bm{R})}{2T}\right]\\
&= \int \dd \omega \nu_{\mathrm{e}}(\bm{R},\omega) \tanh\dfrac{\omega + e\Phi(\bm{R})}{2T}, \label{eq-charge-real-f}
\end{align}
where we use $\nu_{\mathrm{e}}(\bm{R},\omega) = \nu_{\mathrm{h}}(\bm{R},-\omega)$ in the last line.

\begin{figure}
\includegraphics[width = 25em]{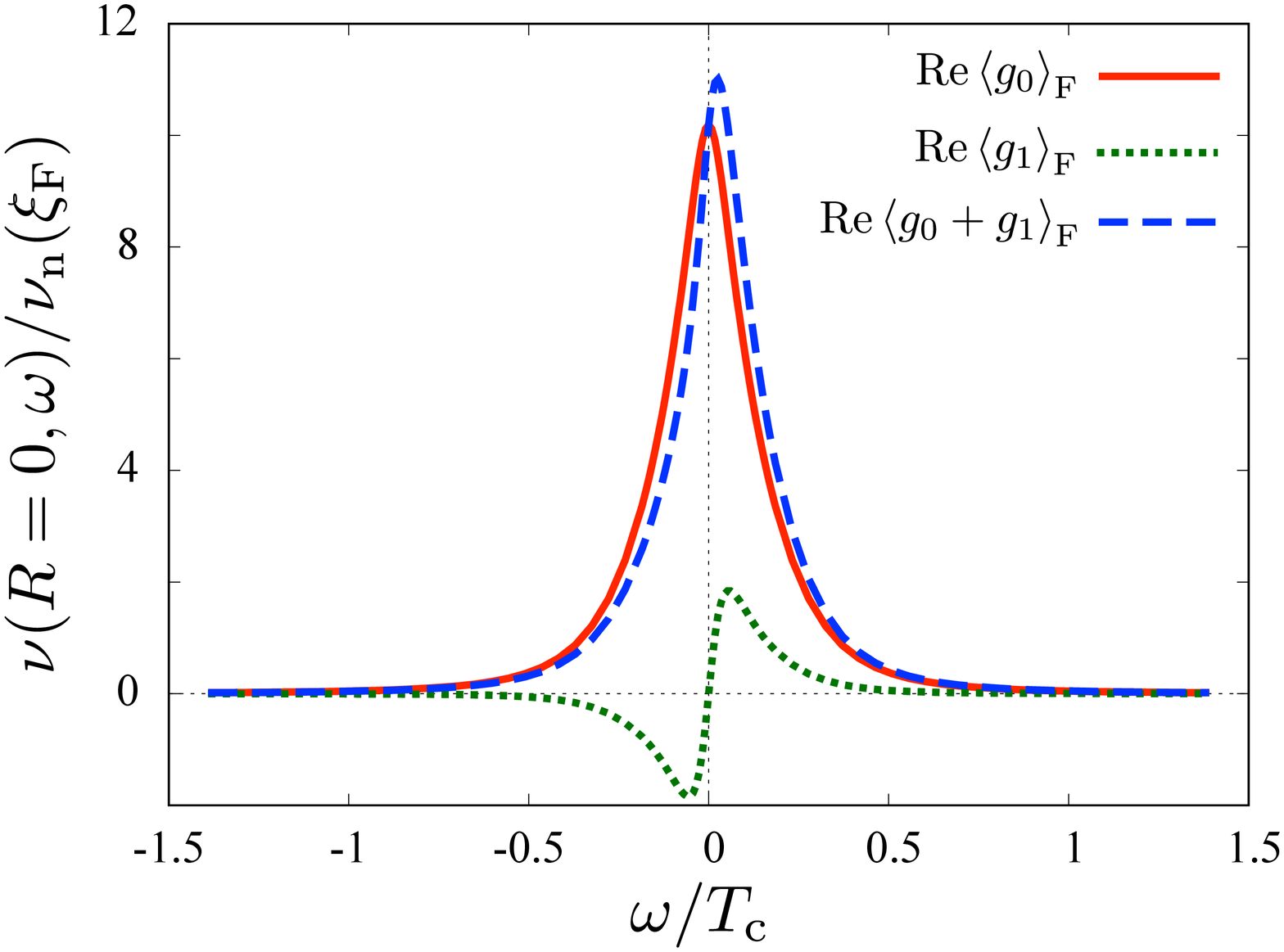}
\caption{Local density of states of an s-wave SC with a single vortex at $R=0$. Here $g_{1}$ is assumed to be $\bar{g}_{1}$, and the parameters are shown in the main text.} 
\label{fig-ldos-s}
\end{figure}

We show asymmetric local density of states $\nu_{\mathrm{e}}(\bm{R},\omega)$. This asymmetry causes the charged vortices, and the remaining origin of the charging effects is the frequency shift by $ e\Phi$ in Eq.~\eqref{eq-charge-real-f}. We use the following parameters: $T=0.4T_{\mathrm{c}}$, $\Gamma_{\n} = 0.05 \Delta_{0}$, $1/(k_{\F}\xi_{0}) =0.01$, and $\lambda_{\mathrm{TF}}/\xi_{0} = 0.01$. We also use the smearing factor to make Green's functions retarded, i.e., $\omega \to \omega + \ii \eta$, which is $0.001T_{\mathrm{c}}$ for $L_{z}= 1$ and $2$, and $0.05 T_{\mathrm{c}}$ for $L_{z} = 0$. This large smearing factor is necessary for finite spectral width of $L_{z} = 0$ because its low energy Andreev bound states hardly broaden against Born-type impurities\cite{Kato2000,JPSJ.71.1721,PhysRevLett.102.117003,Kurosawa2015}. Figure~\ref{fig-ldos-s} shows that the energy spectrum at the origin for $L_{z} = 1$ as an example, and it is clear that the spectral shift should be smaller than the spectral width of the zeroth order quasiclassical Green's function when we use the perturbative ways for the Green's functions. For vanishingly small spectral width, the zeroth order spectrum is given by the delta function and the first order one is proportional to the derivative of the delta function. Hence the perturbative approach does not work and even leads to negative LDOS.

\begin{figure*}
\includegraphics[width = 55em]{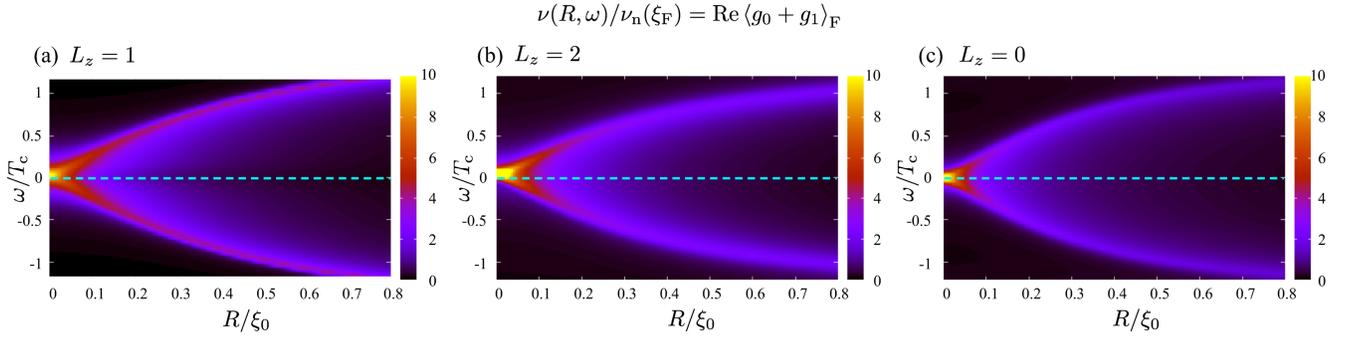}
\caption{Local density of states around the single vortex with (a) $L_{z}=1$, (b) $L_{z}=2$, and (c) $L_{z}=0$. Impurity strength $\Gamma_{\n}/(\pi \Delta_{0}) = 0.05$ for $L_{z}=1$ and $L_{z}=2$, while we utilize a larger smearing factor $\eta/T_{\mathrm{c}} = 0.05$ for $L_{z} = 0$.} 
\label{fig-ldos-all}
\end{figure*}

For those parameters, we show the local densities of states for an s-wave SC and chiral p-wave SCs with a parallel and am antiparallel vortex in Figs.~\ref{fig-ldos-all}(a)-(c), respectively. Spectral asymmetry with respect to $\omega$ appears explicitly for $L_{z} = 1$ and $2$. Slightly larger spectral shift for $L_{z}=2$ than for $L_{z}=1$ is obtained by the following reasons: (i) the larger total angular momentum. (ii) larger Kramer--Pesch effects because of the presence of the induced component and smaller scattering rate than that for $L_{z}=1$ although the scattering rate in the bulk is larger. As we mentioned, the perturbative approach for spectral functions can be guaranteed whether the spectrum has sufficiently large width compared with the spectral shift due to the quantum corrections. Therefore the validity of perturbative approach in a clean system at low temperature for $L_{z} =2$ is somewhat restricted compared with $L_{z}=1$. Although we have introduced the quasiclassical parameter $1/(k_{\F}\xi_{0})$ as a perturbation parameter, it is modified owing to the Kramer--Pesch effects: The core shrinking of the pair potentials causes the larger gradient expansion terms since $1/\xi_{1} > 1/\xi_{0}$. The quantum corrections are larger in lower energy regions and thus the robustness of weakly bound states for $L_{z}=1$ with energy near the superconducting gap does not matter. On the other hand, the perturbative conditions become more severe in chiral p-wave superconducting vortices. For $L_{z} = 0$, the impurity scattering rates are extremely suppressed in the vortex core. In this case, quantitative discussions are difficult because a relatively large smearing factor is introduced, but we have a qualitative feature of no spectral shift at the origin. 

We should remark on the relation between the real frequency formalism and the Matsubara frequency formalism regarding the validity of the perturbative approach.
When we focus on the quasiclassical Green's function describing the Andreev bound states, low Matsubara frequencies at low temperatures stands for the small width in the real frequency formalism. Thus the limitation in the Matsubara formalism in low temperature is related to the validity of the perturbative method in real frequency formalism, in which the validity is  determined by whether the spectral width is sufficiently large at the vortex core.

\section{Summary}
We have investigated the charging effects of single vortices for an s-wave SC and a chiral p-wave SC, and obtained the following results:
(I) We have reproduced the distinct two vortices for the chiral p-wave SC, which was first found using the BdG equation. We have calculated their temperature dependence and found that the distinction is noticeable at low temperatures. In addition, we have clarified that the charging effect for $L_{z} = 2$ is much larger than that for $L_{z} = 1$ for the clean systems. The validity is confirmed by comparing it with numerical results based on the BdG theory. (II) We have also studied the impurity effects on the charge densities. The larger charge for $L_{z}=2$ is strongly suppressed because the energy gap in the bulk is reduced and correspondingly the coherence length becomes longer. This is in contrast to $L_{z}=1$ because of the Anderson's theorem: The small $\Gamma_{\n}$ scatters the core states for $L_{z} = 1$ and reduces the charge density as well as for $L_{z}=2$. The coherence length of the quasiparticles for $L_{z}=1$ becomes shorter for large $\Gamma_{\n}$ in contrast to that for $L_{z}=2$, which determines the spatial scale in the core region and results in recovery of the charge density. (III) The local densities of states have been calculated using the electric field obtained in the Matsubara formalism. We have found that the perturbative approach is valid when the spectral shift is smaller than the spectral width.
After all, our perturbative approach using the augmented quasiclassical theory effectively describes effects of quantum corrections with caution about the validity of the perturbation. Particularly the broadening of the bound-state-spectra caused by such as impurities is necessary, which stands for the difficulty for $L_{z}=0$. 

\section*{Acknowledgements}
The author would like to thank Y.~Tsutsumi and Y.~Kato for helpful discussions. 
This work was supported by a Grant-in-Aid for JSPS Fellows (Grant No.~16J03224) and the Program for Leading Graduate Schools, MEXT, Japan.

\appendix
\section{Force balance in asymptotic region}
\begin{figure}
\centering
\includegraphics[width=25em]{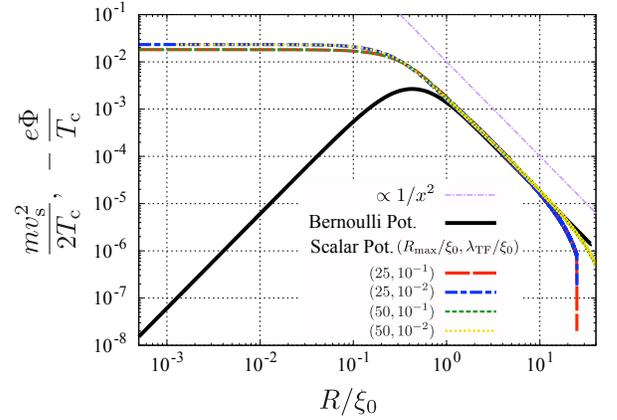}
\caption{Force balance relation. The solid line is the Bernoulli potential. The scalar potential is shown for four set of parameters $(R_{\max}, \lambda_{\TF})$ by dashed lines, where $R_{\max}$ and $\lambda_{\TF}$ represent the system radius used when we solve the Poisson equation and the Thomas--Fermi length. The purple dotted-dashed line shows the behavior of $x^{-2}$ for eye guide.}
\label{fig-force}
\end{figure}
The validity of our calculation can be seen from a force balance relation. 
For simplicity we consider the force balance in the asymptotic region from the vortex core where the London equation holds.
We develop the balance relation from the London equation and the Euler equation by regarding superelectrons as an ideal fluid at $T=0$, which are given by
\cite{London1960,Sonin1998}
\begin{align}
\bm{v}_{\mathrm{s}} &= -\dfrac{e}{mc}\bm{A } + \dfrac{\bm{\nabla}\chi}{2m},\label{eq-london}\\
 \dfrac{\dd \bm{v}_{\mathrm{s}}}{\dd t} &= \dfrac{e}{m}\left[\bm{E} + \bm{v}_{\mathrm{s}} \times \bm{B}\right]. \label{eq-euler}
\end{align}
Equation~\eqref{eq-london} stands for the gauge-invariant velocity of superelectrons using the superconducting phase $\chi$ and $\bm{v}_{\mathrm{s}}$ forms a velocity field. 
We utilize the vector identity $(\bm{v}_{\mathrm{s}}\cdot \bm{\nabla})\bm{v}_{\mathrm{s}} = \bm{\nabla}(v_{\mathrm{s}}^{2}/2) - \bm{v}_{\mathrm{s}}\times (\bm{\nabla}\times \bm{v}_{\mathrm{s}})$ and obtain 
\begin{align}
\dfrac{\partial \bm{v}_{\mathrm{s}}}{\partial t} = \dfrac{e\bm{E} }{m}- \bm{\nabla}\left(\dfrac{v_{\mathrm{s}}^{2}}{2}\right) + \bm{v}_{\mathrm{s}}\times \left[\dfrac{e\bm{B}}{mc}+ \bm{\nabla} \times  \bm{v}_{\mathrm{s}}\right].
\end{align}
The third term gives the delta function resulting from the rotation of the second term in Eq.~\ref{eq-london}, and it does not matter in the asymptotic region.
We consider the equilibrium case, and let the time derivative be zero for the stationary condition. Therefore we derive the balance relation between the Bernoulli potential and the scalar potential:
\begin{align}
\bm{\nabla}\dfrac{mv_{\mathrm{s}}^{2}}{2}-e\bm{E}  = \bm{\nabla}\left[\dfrac{mv_{\mathrm{s}}^{2}}{2} + e\Phi\right] = 0.
\end{align}
For $R\to\infty$, both the Bernoulli potential and the scalar potential are zero, and we find 
\begin{align}
 \dfrac{mv_{\mathrm{s}}^{2}}{2}  =- e\Phi.
\end{align}

Figure~\ref{fig-force} shows these two potentials, which balance with each other in  the asymptotic region $10^{0} \lesssim R/\xi_{0} \lesssim 10^{1}$. 
We set $T=0.4 T_{\mathrm{c}}$ and $\Gamma_{\n} = 0.05\Delta_{0}$. The energy gap is reduced by no more than $2\%$ from $\Delta_{0}$, and the impurity does not affect in the asymptotic region of an s-wave SC. Here we approximately estimate $\bm{v}_{\mathrm{s}} \simeq \bm{j}/(en_{\n})$ with current density $\bm{j} =-e\nu_{\n}(\xi_{\F})\ii \pi T\sum_{n}\braket{\bm{v}_{\F}(g_{0}-\bar{g}_{0})}_{\F}$.
The scalar potential have finite size effects near the outer boundary $R=R_{\max}$.
It also depends on $\lambda_{\TF}/\xi_{0}$ in the core region $R/\xi_{0} \lesssim 10^{-1}$, but effects of $\lambda_{\TF}$ does not appear in the asymptotic region as expected from the above argument.
Since we do not consider the magnetic field, the power-law decay of these two potentials are obtained. The $R$ dependence is given by $R^{-2}$ because 
the velocity induced by the phase winding is $\bm{v}_{\mathrm{s}}\propto \bm{\nabla}\chi = \hat{e}_{\theta}R $ for $\chi = \theta$. We also show the behavior of $1/(R/\xi_{0})^{2}$, which agrees with the decay of the two potentials. Therefore we conclude that the charge density obtained by our perturbative approach is consistent with the physical picture obtained by the simple argument on the force balance relation in the asymptotic region.

\bibliographystyle{apsrev4-1}
\bibliography{library}

\end{document}